# The Economic Dividends of Peace: Evidence from Arab-Israeli Normalization[1]

Mitja Kovac          Rok Spruk

**Abstract**

*This paper provides the first causal evidence on the long-run economic dividends of Arab-Israeli peace treaties. Using synthetic control and difference-in-differences estimators, we analyze Egypt's 1978 Camp David Accords and Jordan's 1994 peace treaty with Israel. Both cases reveal large and lasting gains. By 2011, Egypt's real GDP exceeded its synthetic counterfactual by 64 percent, and per capita income by 82 percent. Jordan's trajectory shows similarly permanent improvements, with real GDP higher by 75 percent and per capita income by more than 20 percent. The mechanisms differ: Egypt's gains stem from a sharp fiscal reallocation together with higher foreign direct investment and improved institutional credibility, while Jordan benefited primarily through enhanced trade and financial inflows. Robustness and placebo tests confirm the uniqueness of these effects. The results demonstrate that peace agreements yield large, durable, and heterogeneous growth dividends.*

---

[1] Kovac: Professor of Law and Economics, School of Economics and Business, University of Ljubljana, Kardeljeva ploscad 17, SI-1000 Ljubljana. E: mitja.kovac@ef.uni-lj.si. Spruk: Associate Professor of Economics, School of Economics and Business, University of Ljubljana, Kardeljeva ploscad 17, SI-1000 Ljubljana. E: rok.spruk@ef.uni-lj.si

**Introduction**

Armed conflict and peace treaties are among the most consequential institutional shocks in modern political economy. While a large body of research has documented the destructive effects of conflict on economic growth, development, and institutional quality (Barro 1991, Collier 1999, Collier and Hoeffler 2004, Abadie and Gardeazabal 2003, Miguel and Roland 2011), far less is known about the long-term economic dividends of peace. The notion of a "peace dividend", that ending hostilities and normalizing relations can unlock growth, investment, and welfare gains, has long animated public discourse and policymaking, but systematic quantitative evidence remains scarce. This paper provides one of the first causal estimates of the economic benefits of peace in the context of the Arab-Israeli conflict.

We focus on the 1978 Camp David Accords, in which Egypt became the first Arab country to sign a formal peace treaty with Israel. The agreement not only transformed the strategic balance of the Middle East but also redefined Egypt's economic trajectory by reducing military burdens, facilitating U.S. aid, and opening avenues for trade and investment. To identify the causal impact of peace on long-run economic performance, we apply the synthetic control method (Abadie and Gardeazabal 2003, Abadie, et. al., 2010, 2015), constructing a counterfactual growth path for Egypt in the absence of peace treaty. The donor pool consists of states that have never normalized relations with Israel from 1950 onward, allowing us to benchmark Egypt's trajectory against a credible set of "non-normalizers."

This analysis builds on and contributes to several strands of literature. First, it extends the empirical literature on the economic consequences of war and conflict termination (Miguel et. al. 2004, Besley and Persson 2008, Blattman and Miguel 2010), which has largely emphasized the costs of violence rather than the benefits of peace. Second, it contributes to the literature on international institutions and cooperation (Keohane 1984, Fearon 1995, Mansfield and Pevehouse 2000), by showing how peace agreements function as institutional commitments with measurable developmental returns. Third, it speaks to recent work using synthetic control methods to study political and institutional shocks (Billmeier and Nannicini 2013, Pinotti 2015, Cavallo et al. 2013, Abadie 2021), situating the Arab-Israeli peace process within the broader toolkit of comparative political economy.

Our findings reveal substantial long-term economic gains from peace. Egypt's post-1978 trajectory diverges markedly from its synthetic counterfactual, with large and persistent improvements in GDP per capita, investment, and structural transformation. These results suggest that the peace treaty not only ended a costly cycle of conflict but also enabled a fundamental reorientation of Egypt's economic development. The evidence is consistent with both reduced military spending (Dunne et. al. 2005), increased foreign assistance (Burnside and Dollar 2000), and enhanced trade and capital inflows (Rose 2004, Subramanian and Wei 2007) as mechanisms linking peace to prosperity.

Beyond historical interest, the results have pressing contemporary implications. The recent wave of Arab-Israeli normalization agreements, including Jordan's 1994 treaty, the Abraham Accords of 2020 (UAE, Bahrain, Morocco, and Sudan), and ongoing discussions of a Saudi-Israeli agreement, raises the question of whether peace continues to generate tangible economic dividends. By providing rigorous evidence from the foundational case of Egypt, this paper informs both scholarly debates and policy discussions on the economic consequences of peace in the Middle East and beyond.

The rest of the paper is organized as follows- The first part (Section 2) offers an overview of previous literature. The second part (Section 3) provides a detailed discussion on the mechanisms at work, whereas the third part (Section 4) discusses historical, legal and political background. Fourth part (Section 5) offers description of the data and of the sample design, the sixth part (Section 6) thoroughly discusses identification strategy and seventh part (Section 7) examines results and their implications. Eight part (Section 8) concludes.

## 2   Previous Literature

The relationship between armed conflict, peace agreements, and economic development has long been debated in both economics and political science. Scholars have converged on the view that violent conflict imposes large and persistent costs on economies, yet systematic evidence on the benefits of peace remains surprisingly limited. At the same time, advances in causal inference have made it increasingly feasible to evaluate the developmental consequences of institutional shocks such as peace treaties. This section reviews the relevant bodies of work on conflict and development, peace dividends, international cooperation, and methodological innovations.

*Conflict and development.* A substantial literature documents the destructive effects of armed conflict on economic outcomes. Early cross-national analyses established robust associations between civil wars and diminished growth, weaker institutions, and delayed development (Collier 1999, Collier and Hoeffler 2004, Gates et. al. 2012). More disaggregated studies suggest that the costs are not confined to aggregate GDP but extend to poverty, employment, and human capital (Verwimp et. al. 2009; Bozzoli and Brück 2009). Recent contributions demonstrate that the scars of conflict are long-lasting. For instance, Costalli et. al. (2017) show persistent macroeconomic effects of armed conflicts, while Galindo-Silva and Tchuente (2023) document sharp declines in educational outcomes during Cameroon's Anglophone conflict. These findings suggest that conflict fundamentally alters development trajectories, raising the question of whether peace can provide a durable counterweight.

*Peace dividends and recovery.* Despite extensive work on conflict costs, rigorous analysis of peace dividends remains limited. Policymakers have often invoked the term "peace dividend" to capture the expectation that military demobilization and reduced defense spending would stimulate growth (Knight et. al. 1996, Dunne et. al. 2005). Empirical studies point to increased aid, trade, and growth opportunities following peace agreements, particularly in post-conflict African states (Collier and Hoeffler 2002, Flores and Nooruddin 2009). Yet the evidence is mixed. By way of example, Fajardo-Steinhäuser (2023) finds limited short-run economic gains from Colombia's peace agreement despite major reductions in violence. More broadly, Ghobarah et. al. (2003) show that the human costs of war persist long after hostilities end, complicating the expectation of immediate dividends. Taken together, the literature emphasizes that while peace may create conditions for recovery, its economic impact is context-dependent and not well understood in interstate settings.

*International cooperation and institutions.* A related body of work examines how treaties and institutions alter incentives and promote economic integration. Research in political economy and international relations emphasizes the importance of credible commitments for sustaining cooperation (Keohane 1984, Fearon 1995, Fortna 2004). Empirical studies show that preferential trade agreements and membership in international organizations increase trade and reduce conflict risk (Mansfield and Bronson 1997, Mansfield and Pevehouse 2000, Martin et. al. 2008). However, peace treaties, arguably the most fundamental institutional agreements, have received limited attention as economic events. The Arab-Israeli agreements, in particular, are typically studied through security or diplomatic lenses rather than developmental ones.

*Methodological advances.* Recent methodological innovations make it possible to credibly estimate the effects of rare, large-scale political events. The synthetic control method has been applied to diverse contexts, including terrorism, democratization, and natural disasters (Costalli et al. 2017, Billmeier and Nannicini, 2013, Cavallo et. al. 2013). Extensions and refinements have further expanded its applicability (Xu 2017, Dube and Zipperer 2015, Galiani and Quistorff 2017). These developments provide a powerful toolkit for analyzing single-country interventions such as peace treaties, where randomized experiments are impossible and conventional panel estimators often fail.

In summary, the literature has convincingly established the costs of conflict and provided suggestive evidence of peace dividends, yet the causal impact of interstate peace treaties on long-run economic development remains underexplored. Our paper addresses this gap by applying synthetic control methods to the case of Egypt's 1978 Camp David Accords, thereby offering the first rigorous counterfactual analysis of the economic benefits of Arab-Israeli peace.

## 3     Mechanisms at Work

Understanding the economic consequences of peace requires identifying the channels through which a peace agreement may alter a country's development trajectory. In principle, peace may be little more than the absence of conflict, with no major economic implications. Yet theories of political economy and international relations suggest that peace treaties can be institutional shocks that reshape fiscal choices, international relationships, and credibility in ways that produce lasting developmental gains. Establishing these channels is crucial, not only to interpret our empirical results but also to assess whether the benefits of peace are likely to generalize across contexts.

The existing literature highlights four broad mechanisms. First, peace reduces military expenditures and thereby frees resources for productive investment. Second, peace often triggers inflows of foreign aid and external resources, as donor governments and institutions reward stabilization. Third, peace lowers political risk and creates opportunities for expanded trade and foreign direct investment. Fourth, peace can strengthen institutional credibility by signaling long-term stability and reducing uncertainty. Each mechanism is rooted in theoretical arguments and supported by empirical evidence from leading studies in economics and political science.

### 3.1     *Defense Burden and Fiscal Constraints*

The most immediate effect of peace is the reduction of military expenditures. High defense spending is consistently associated with crowding out productive investment (Knight et. al. 1996, Dunne et. al. 2005). Gupta et. al. (2004) show that military spending is negatively correlated with growth in developing countries, largely by diverting resources from infrastructure and social spending. More broadly, Barro and Sala-i-Martin (2004) emphasize that fiscal space is a key determinant of long-term development. Peace therefore expands the fiscal room for investments in human capital, health, and physical infrastructure that are central to sustained growth.

### 3.2  External Transfers and Aid Flow

Peace treaties often catalyze large inflows of external resources. Burnside and Dollar (2000) show that aid effectiveness is conditional on institutional quality, while Collier and Hoeffler (2002) highlight the exceptional growth opportunities in post-conflict contexts. Clemens et. al. (2012) further refine the debate by distinguishing between short- and long-term aid effects, finding robust positive effects in the short run. Nunn and Qian (2014) provide evidence on how aid can influence development through trade linkages, reinforcing the notion that external transfers may play a transformative role when combined with stability. Egypt's experience after 1978 illustrates this dynamic: U.S. assistance expanded dramatically, providing both fiscal relief and foreign exchange to support investment.

### 3.3  Trade, Investment and Economic Integration

Peace can improve access to markets by lowering political risk. Rose (2004) and Subramanian and Wei (2007) show that trade institutions substantially boost international flows. Furthermore, Mansfield and Pevehouse (2000) and Mansfield and Reinhardt (2008) emphasize that formal commitments through treaties reduce uncertainty and promote both political and economic cooperation. In addition, Martin et. al. (2008) argue that trade can reduce the probability of conflict, highlighting a feedback loop between peace and integration. Evidence from globalization studies confirms that openness is linked to faster growth (Frankel and Romer, 1999). Similarly, Alfaro et. al. (2008) demonstrate that institutions strongly condition foreign direct investment flows. In this sense, peace agreements can operate much like trade agreements by lowering risk, improving credibility, and attracting investment.

### 3.4  Institutional Credibility and Expectations

Finally, peace treaties may function as credibility-enhancing devices. Fearon (1995) highlights the role of credible commitments in avoiding conflict, while Keohane (1984) emphasizes the importance of institutional foundations of cooperation. Empirical evidence strongly links institutional stability to economic development. For example, Acemoglu et. al. (2001) show that historical institutions shape long-run growth, and Acemoglu et. al. (2019) confirm that democracy boosts economic development. North (1990) provides the theoretical foundation for understanding institutions as constraints on uncertainty. More recently, Besley and Persson (2011) highlight state capacity as critical to both conflict prevention and development. Peace agreements, by committing states to stability, may thus enhance credibility, reduce risk premiums, and thereby stimulate long-term investment.

The literature therefore points to four channels through which peace can generate economic dividends, namely, reduced military burdens, expanded external transfers, deeper integration into trade and investment networks, and stronger institutional credibility. Each mechanism has strong theoretical and empirical foundations in the top economics and political science literatures. The empirical contribution of this paper is to evaluate the aggregate effect of these channels by constructing a credible counterfactual for Egypt's economic trajectory absent the 1978 peace treaty.

## 4     Historical and Policy Background

### 4.1     *Arab-Israeli Conflict: The Context*

The origins of the Camp David Accords lie in three decades of recurrent wars between Israel and its Arab neighbors following the creation of Israel in 1948 (Maoz 2006). The 1948-49 Arab-Israeli war, the Suez Crisis of 1956, and most decisively the Six-Day War of 1967 left Israel in control of significant territories, including the Sinai Peninsula, the West Bank, Gaza, and the Golan Heights (Shlaim 2000; Stein 1989; Filiu, 2014). For Egypt, the loss of the Sinai not only carried strategic and territorial significance but also created enormous fiscal and military pressures (Podeh 1999; Lenczowski, 1990). Under President Gamal Abdel Nasser and, after 1970, Anwar Sadat, Egypt struggled with the economic consequences of repeated conflict, stagnating growth, and rising social discontent (Hinnebusch 1985; Quandt, 1986). By the mid-1970s, Sadat increasingly sought a strategic reorientation: disengaging from the Soviet Union, pursuing economic liberalization through the infitah, and exploring peace with Israel as a pathway to recovering Sinai, reducing military expenditures, and securing access to Western aid and investment (Sayigh 1991; Khalidi, 2013). U.S. diplomacy under Presidents Nixon, Ford, and Carter provided fertile ground, with Carter in particular determined to pursue a comprehensive Middle East settlement (Quandt 1988, 2005; Filiu, 2014, Avraham, 2002). Sadat's dramatic 1977 visit to Jerusalem, the first by an Arab leader, marked a turning point by breaking long-standing taboos and opening the path to bilateral negotiations under U.S. mediation.

### 4.2 The Camp David Accords and the Egypt-Israel Peace Treaty (1978-79)

The Camp David Accords, signed in September 1978 by Anwar Sadat, Israeli Prime Minister Menachem Begin, and U.S. President Jimmy Carter, represented a diplomatic breakthrough of historic proportions (Stein 1999; Brzezinski, 1983). The accords contained two frameworks: one outlining a general vision for peace in the Middle East, including provisions for Palestinian autonomy in Gaza and the West Bank, and the other establishing the basis for a bilateral treaty between Egypt and Israel (Quandt, 2011). The subsequent treaty, signed in March 1979, entailed Israel's complete withdrawal from the Sinai Peninsula in exchange for Egypt's formal recognition of Israel and the establishment of diplomatic relations (Telhami 1992; Ashton, 2017; Brams and Togman, 1996). The United States provided extensive guarantees, including significant economic and military aid packages to Egypt (Sharp 2005).

The consequences for Egypt were immediate. The treaty secured the recovery of Sinai and ended recurrent wars with Israel, but also provoked backlash across the Arab world. Egypt was expelled from the Arab League, a suspension that lasted until 1989 (Podeh 2015). Despite regional isolation, Egyptian leaders judged that the strategic and economic benefits outweighed the political costs. Large inflows of U.S. assistance began to flow, military expenditures as a share of GDP began to fall, and Egypt realigned firmly with the West in the Cold War order (Rubin 1994).

### 4.3 Later Normalization Efforts: The Abraham Accords and Beyond

Egypt's treaty was followed by Jordan's 1994 peace treaty with Israel, and more recently by the Abraham Accords. Signed in 2020, the accords brought Israel into normalized relations with the United Arab Emirates, Bahrain, and later Morocco, with Sudan also pledging normalization (Miller 2020). Unlike Egypt's case, these agreements were not preceded by direct interstate wars but were instead motivated by shifting geopolitical dynamics, particularly shared concerns over Iranian influence and a desire to cement closer ties with the United States (Lynch 2021).

The Abraham Accords also differed in economic context. The Palestinian issue, central to the Egyptian-Israeli negotiations, was less salient in the Gulf agreements, where normalization was decoupled from immediate resolution of the Israeli-Palestinian conflict (Guzansky and Marshall 2020). Economic incentives played a central role: the Gulf states sought diversification away from oil and saw in Israel a partner in technology, finance, and security cooperation (Cohen 2021). Thus, whereas Egypt's treaty was largely about recovering territory and unlocking aid, the Abraham Accords were forward-looking in economic integration and modernization.

*4.4 Arab Opposition to the Recognition of Israel and its Legitimacy*

The Arab-Israeli conflict has been a self-reinforcing conflict since its earliest days, characterized by terrorism and an "us or them" mentality that has worked to dehumanize the parties in each other's eyes (Horowitz 2024).The questioning of Israel's legitimacy (and related non-recognition of Israel by several states that form also our donor pool) has in fact accompanied the Jewish state from its very foundation as a by-product of the larger Arab-Israeli conflict (Barzilai, 2015). The sources of this early wave of delegitimization were the Arab states which evoked for years through their state-controlled media the image of Israel as being created by a colonial-settler movement, backed by Western imperialism, with no authentic connection to the land which it claimed (Gold, 2010; Taulbee and Forsythe, 1972). From the beginning of Egyptian reconciliation with Israel and particularly Sadat's visit to Israel stirred many angry reactions in the Arab world as a whole and these found expression in noisy demonstrations that took place in various Arab countries and cities, especially Iraq, Iran, Syria, Tunisia, Libya and Algeria (Akram and Lynk, 2013). The stiff opposition towards recognition of Israel as an independent state in countries that form our donor pool countries derives primarily from ideological stands, fears for Arab interests, and above all an assessment of the current international, Arab and Israeli situation that differed radically from Sadat's (Egyptian) or Jordanian assessment and argued that the visit to Israel and consequential peace and recognition of Israel would only strengthen the position of the hard-line right-wing Likud government (Akram and lynk, 2013; Jiryis, 1978).

Iraq for example condemned Egypt for concluding peace with Israel and attacked Egypt for going along Israel by following the path of a political settlement (Jiryis, 1978). Iraq has, for some time, been also calling for a condemnation of UN Security Council resolutions 242 and 338, the rejection of the Geneva Conference, and the use of force as a way of liberating the occupied territories (Jiryis, 1978). In Libya, the General People's Congress issued a statement condemning Sadat's to visit the occupied land (Israel) and to hold discussions with the terrorist Menahem Begin (Akram and Lynk, 2011, 2013). Libyan Congress also declared that Sadat's visit and peace with Israel constituted a crime against the entire Arab nation which it cannot ignore or be silent about (Jiryis, 1978). In addition, Libyan government had taken severe economic and political measures against the Egyptian government (Wright, 1988). For decades the Arab nations (with few exemptions) have continued to assert the claim of the Palestinian Arabs to Palestine, rejecting both the Balfour Declaration and the United Nations Partition Resolution (Akram and Lyink, 2011; Schueller, 1978). The Israelis, on the other hand, point to these documents as legitimate sources of international territorial sovereignty, the legitimacy of which is reinforced by Israel's recognition as a member state in the United Nations (Schueller, 1978).

The positions towards peace with Israel adopted by most other Arab countries (Syria, Sudan, Algeria, South Yemen, Tunisia, Lebanon, Saudi Arabia) did not differ much in essentials from those outlined above although they were cast in more moderate tones (Wright, 1968; Jiryis, 1978). Thus, since the creation of Israel in 1948, most Arab states have refused to engage in official diplomatic relations with it and although Egypt and Jordan signed formal peace agreements with Israel, many Arab states continued their boycott (Hallward and Biyagutane, 2024). Finally, in 2020, the United Arab Emirates (UAE), Bahrain, Sudan, and Morocco recognized Israel and established diplomatic relations under the US-sponsored Abraham Accords framework (Hallward and Biyagutane, 2024).

### 4.5     Policy Conditions Enabling Normalization and Peace

In both Camp David and the Abraham Accords, several enabling conditions stand out. U.S. mediation and guarantees were pivotal, both in Carter's role at Camp David and in the Trump administration's sponsorship of the Abraham Accords (Quandt 2005, Miller 2020). Strategic threats provided urgency. For Egypt, the repeated wars with Israel were unsustainable. For the Gulf, Iranian influence and regional instability played a similar role. Domestic leadership was decisive. Sadat's willingness to pursue peace despite domestic opposition and regional isolation echoes the political calculations of Gulf leaders who embraced normalization despite ideological costs (Telhami 1992, Lynch 2021). Finally, economic imperatives were central. Egypt faced stagnation, fiscal stress, and a growing population; the Gulf states today face the challenge of oil dependence and seek diversification (Cohen 2021).

### 4.6     Implications for Counterfactual Analysis

The historical record has direct implications for empirical analysis. Egypt's Camp David treaty was a watershed involving territorial restitution, demilitarization, and massive U.S. aid flows. It thus represents a suitable setting for evaluating the long-run economic consequences of peace. Later agreements, such as the Abraham Accords, are less comprehensive and more recent, making it harder to identify durable long-term effects, but they underscore how peace can quickly expand trade, tourism, and investment flows. Moreover, global context also matters. Egypt's 1978 agreement occurred in a Cold War environment where U.S. aid and security alignment were central, whereas today's normalization agreements occur in a globalized economy where trade, technology, and finance dominate. These differences earmark both the uniqueness of the Egyptian case and its usefulness for identifying the economic dividends of peace.

## 5 Data and Sample Design

The analysis relies on a novel dataset that combines long-run national accounts, sectoral aggregates, and institutional indicators, harmonized across countries and across time. The guiding principle of the design is to capture both the aggregate economic consequences of Arab-Israeli normalization and the fiscal, financial, and institutional mechanisms through which such effects operate. In this sense, the dataset follows the tradition of integrating synthetic control methods with historically grounded macroeconomic series, as pioneered by Abadie et. al. (2010), and developed further in applications of institutional shocks and peace agreements (Acemoglu et. al. 2008, Besley and Persson 2009).

### 5.1 Aggregate outcomes

The primary dependent variable is real GDP, expressed in PPP-adjusted international dollars at constant prices, taken from Feenstra et. al (2015). These authors provide harmonized output-side real GDP measures ensuring international comparability across both space and time. To capture welfare at the individual level, we use real GDP per capita from the updated Maddison Project Database (Bolt and van Zanden 2014), which offers superior historical depth by extending coverage into the postwar period. The combination of Feenstra et al. (2015) and updated Maddison database ensures that the analysis benefits from both consistency in cross-country methodology and the historical coverage required for evaluating long-run effects of treaties signed in 1978 and 1994.

### 5.2 Transmission mechanism

Four additional outcome measures capture the key channels emphasized in the literature on the peace dividend and post-conflict growth (Collier et. al. 2008, Gates et al. 2012). First, military expenditure as a share of GDP comes from the *Stockholm International Peace Research Institute* (SIPRI). This series, widely regarded as the benchmark in defense economics, offers the most reliable time-series evidence on demobilization and fiscal reallocation. Although SIPRI notes limitations in early coverage, its comparability and long-run continuity outweigh these concerns, making it the standard measure for identifying fiscal peace dividends.

Second, foreign direct investment inflows are measured as a share of GDP, obtained from the World Development Indicators (WDI). The WDI, drawing on IMF Balance of Payments Statistics, provides the longest consistent coverage of annual FDI inflows across the region. We focus on inflows rather than stocks because flows capture investors' contemporaneous responses to geopolitical risk, credibility of treaties, and integration into international capital markets (Jensen 2008, Alfaro et. al. 2004,).

Third, trade openness is defined as the share of exports plus imports in GDP, based on Feenstra et al. (2015). This indicator has become the canonical measure in international economics (Frankel and Romer 1999, Rodrik 1999) and has been widely adopted to study globalization, trade shocks, and conflict. The measure directly reflects external economic integration relative to domestic production and is well suited for capturing whether peace translates into expanded market access.

Fourth, private household consumption as a share of GDP is taken from Feenstra et. al. (2015). Although U.S. bilateral aid inflows would constitute a direct channel of fiscal transfers linked to peace agreements, historical aid coverage is incomplete, particularly prior to the 1990s. Household consumption shares therefore provide the most comprehensive and consistently available indicator of potential welfare spillovers. In the context of U.S. aid to Egypt and Jordan, consumption shares capture the indirect effects of foreign transfers that relax budget constraints and expand domestic welfare. While aggregate consumption is an imperfect proxy, the advantages of comparability and long-run coverage outweigh its limitations.

### 5.3 Institutional mechanisms

To capture institutional change, we rely on the Varieties of Democracy (V-Dem) dataset (Coppedge et al. 2021). V-Dem offers the most comprehensive and historically extensive set of institutional indicators currently available. We focus on three theoretically grounded clusters. The first cluster concerns judicial institutions and the rule of law, including judicial independence, compliance with judiciary, and equality before the law. These measures capture the credibility of courts and impartiality of law enforcement, both of which are central to investor confidence and the ability of governments to commit credibly (La Porta et. al. 2008, North et. al. 2009).

The second cluster relates to executive integrity, including executive respect for the constitution, executive corruption, and executive integrity. These variables allow us to examine whether peace agreements alter the constraints on executive behavior and reduce rent extraction, a key prediction of the credible commitment literature (Fearon 1998, Acemoglu and Robinson 2012).

The final cluster captures civil society and accountability, measured by civil society strength. This reflects the monitoring and disciplining capacity of non-state actors and is theoretically relevant in post-conflict environments where external actors provide enforcement guarantees (Przeworski 2005). While perception-based, these V-Dem measures remain the gold standard in comparative institutional analysis and offer the best combination of comparability, historical depth, and methodological transparency.

### 5.4 Sample design

The treated units are Egypt, which signed the Camp David Accords in 1978, and Jordan, which signed the Israel-Jordan Peace Treaty in 1994. These two cases represent the first and second waves of Arab-Israeli normalization, constituting major geopolitical shocks to long-standing conflictual equilibria. The donor pool consists of countries without diplomatic ties or official recognition of Israel from 1960 onward: Algeria, Bangladesh, Cuba, Indonesia, Iran, Iraq, Kuwait, Lebanon, Libya, Malaysia, Nicaragua, Oman, Pakistan, Qatar, Saudi Arabia, Syria, Tunisia, and Yemen. By design, this ensures that the counterfactual is not contaminated by contemporaneous normalization processes. The donor group is also regionally relevant and shares similar structural characteristics with the treated units.

This combination of dependent variables, mechanism indicators, and institutional measures provides an unusually rich basis for examining both the economic consequences of normalization and the channels through which peace treaties translate into growth. By combining output measures, fiscal and financial aggregates, and institutional indicators, the dataset allows us to disentangle competing explanations and link the Arab-Israeli case to broader debates on peace dividends, credible commitments, and institutional development.

# 6 Identification Strategy

## 6.1 Setup

Our goal is to examine the contribution of peace and normalization with Israel to economic performance and institutional quality in the Middle East. To this end, our aim is to estimate the appropriate counterfactual growth and governance trajectories to elicit the economic and institutional costs of remaining in a state of unresolved conflict.

Suppose we observe a finite set of countries $(J + 1) \in \mathbb{N}$ over $T \in \mathbb{N}$ periods where $t = 1, 2 \ldots T$. A normalization process that entails the characteristics of the treatment-based policy shock occurs at time $T_0$ and begins in $T_0 + 1$ and lasts until the end of the time period without interruption so that $t < T_0 < T$ and $T_0 \in \{1, T\} \cap \mathbb{N}$ (Abadie 2021). Let $q_{j,t}^N$ be the potential economic and institutional outcome in j-th region in the hypothetical absence of the normalization and peace treaty for $j \epsilon \{1, \ldots J + 1\}$ and $t \in \{1, \ldots T\}$, and let $q_{j,t}^I$ represent the observed realization of the outcome in the full period $t = 1, 2, \ldots T$. Without the loss of generality, the average effect of the peace treaty is defined as:

$$\alpha_{j,t} = q_{j,t}^I - q_{j,t}^N \tag{1}$$

where $\alpha_{j,t}$ captures the difference between the observed realization of the outcome and the potential outcome in the hypothetical absence of the normalization and peace treaty. Since the peace can be operationalized as a binary variable switching between 0 and 1 and sub-periods $t < T_0$ and $t \geq T_0$, policy treatment variable can be described as a dummy variable that takes the value $D_{j,t} = 1 \ \forall \ \{j = 1\}$ and $D_{j,t} = 0$ otherwise. The effect of the peace treaty can be written as follows:

$$q_{j,t} = q_{j,t}^N + \alpha_{j,t} \cdot D_{j,t} \tag{2}$$

The major identification constraint in estimating $\alpha_{j,t}$ arises from the fact that $q_{j,t}^N$ is unobserved to the econometrician and therefore has to be estimated to gauge the effect of the peace treaty. In this respect, our aim is to estimate the full vector of post-treatment effects of normalization process $(\alpha_{1,T_0}, \ldots \alpha_{1,T})$. This, let $q_{j,t}^N$ be approximated through a simple latent factor model:

$$q_{j,t}^N = \delta_t + \boldsymbol{\theta}_t \cdot \mathbf{Z}_{j \in J} + \boldsymbol{\lambda}_t \cdot \boldsymbol{\mu}_{j \in J} + \varepsilon_{j,t} \tag{8}$$

where $\delta_t$ denotes the full set of time-fixed effects that absorb time-varying technology shocks common to all regions with constant loading, $\mathbf{Z}$ is a simple $(m \times 1)$ vector of observed auxiliary covariates unaffected by the peace treaty, $\boldsymbol{\theta}_t$ is $(1 \times m)$ vector of prior unknown parameters, $\boldsymbol{\lambda}_t$ is $(1 \times H)$ vector of observed common factors, and $\boldsymbol{\mu}_j$ is $(H \times 1)$ a vector of unknown factor loadings, and $\varepsilon$ is the set of country-level transitory shocks under $\varepsilon \sim i.i.d$ structure (Xu 2017)

As noted by Firpo and Possebom (2018), let $\mathbf{Y}_{j,t} = [Y_{j,1}, \ldots Y_{T_0}]$ be a vector of the observed realization of the outcomes for county $j \in \{1, \ldots J+1\}$ in the pre-intervention period where $t < T_0$, and let $\mathbf{X}_{j,t} = [X_{j,1}, \ldots X_{T_0}]$ be $[K \times 1]$ vector of covariates. Moreover, let $\mathbf{Y}_{j,t} = [\mathbf{Y}_2, \ldots \mathbf{Y}_{J+1}]$ be a matrix with $[T_0 \times J]$ dimension, and $\mathbf{X}_{j,t} = [\mathbf{X}_2, \ldots \mathbf{X}_{J+1}]$ a corresponding $[K \times J]$. Moreover, let $\mathbf{W} = (w_1, \ldots w_{J+1})$ be a simple vector of weights with $J \times 1$ dimension that captures the composition of the donor pool. Each particular value of the weight vector captures the weighted average of the donor pool's characteristics that best reproduce the economic and institutional trajectory prior to the institutional integration implied from $\mathbf{Y}_{j,t}$ and $\mathbf{X}_{j,t}$ matrices. Thus, the fitted value of the outcome variable constructed from the characteristics of the donor pool for the given $\mathbf{W}$ is as follows:

$$\sum_{j=2}^{J+1} w_j \cdot q_{j,t} = \delta_t + \theta_t \cdot \sum_{j=2}^{J+1} w_j \cdot Z_j + \lambda_t \cdot \sum_{j=2}^{J+1} w_j \cdot \mu_j + \sum_{j=2}^{J+1} w_j \cdot \varepsilon_{j,t} \tag{3}$$

which implies that for each $t \in \{1, \ldots T\}$, we estimate $\hat{q}_{i,t}^N = \sum_{j=2}^{J+1} \hat{w}_j \cdot q_{j,t}$. The optimal weights are derived by dividing the pre-normalization period training and validation sub-periods. which consists of the training and validation period. In the training period, the relative importance of covariates and pre-normalization outcomes is identified through a diagonal matrix $\hat{\mathbf{V}}$ denoting the normalized variable weights. In the validation period, the weighing vector $\hat{\mathbf{W}} = [\hat{w}_1, \ldots \hat{w}_{J+1}] \in \mathbb{R}^J : w_j \geq 0$ for each $j \in \{2, \ldots J+1\}$ captures the relative importance of each region in the loci of treated country's convex hull (Botosaru and Ferman 2019) where the set of weights selected on the basis of the similarity between $J = 1$ and $j \in \{2, \ldots J+1\}$ is given as a closed-form solution to the nested minimization problem (Becker and Klößner 2018):

$$\hat{\mathbf{W}}(\mathbf{V}) = \underset{\mathbf{W} \in \mathbb{W}}{\operatorname{argmin}} (\mathbf{X}_1 - \mathbf{X}_0 \mathbf{W})' \mathbf{V} (\mathbf{X}_1 - \mathbf{X}_0 \mathbf{W}) \tag{4}$$

where $\mathbb{W} = \left\{ W = [w_2 \ldots w_{J+1}]' \in \mathbb{R}^J : w_J \geq 0 \text{ for each } j \in \{2, \ldots J+1\} \text{ and } \sum_{j=2}^{J+1} w_j = 1 \right\}$ and **V** is a diagonal positive semi-definite matrix having $K \times K$ dimension with a trace equal to one:

$$\mathbf{V} = \underset{\mathbf{V} \in \mathbb{V}}{\mathrm{argmin}} \left( \mathbf{X}_1 - \mathbf{X}_0 \widehat{\mathbf{W}}(\mathbf{V}) \right)' \left( \mathbf{X}_1 - \mathbf{X}_0 \widehat{\mathbf{W}}(\mathbf{V}) \right) \tag{5}$$

where notice that **W** is a weighing vector measuring the relative importance of weights from $j \in \{2, \ldots J+1\}$ potential sequence in the donor pool in the composition of synthetic control group, and **V** measures the relative predictive importance of each of K covariates and pre-$T_0$ outcomes (Billmeier and Nannicini 2013). Both weighing vector and covariate-level vector track and reproduce the economic and institutional trajectories of Egypt and Jordan as closely as possible. Hence, by choosing an appropriate distance matrix such as Euclidean or Trigonometric, the relative discrepancy between the treated region and its synthetic control group can be evaluated accordingly (Doudchenko and Imbens 2016). Without the loss of generality, the treatment effect of the peace treaty on economic and institutional outcome for $J = 1$ country and each $t \in \{1, \ldots T\}$ can be written as:

$$\lambda_{J=1,t} = q_{1t} - \sum_{j=2}^{J+1} w_j^* \cdot q_{j,t} = q_{1t} - \hat{q}_{1t}^N \tag{6}$$

### 6.2 Inference

To evaluate the significance of the effect of peace with Israel, our approach relies on the standard permutation test. In this respect, the question we ask is whether the estimated institutional and economic outcome gap is obtained by chance. Following Abadie and Gardeazabal (2003), Bertrand et. al. (2004), Abadie et. la. (2010), McClelland and Gault (2017), Firpo and Possebom (2018), we perform a series of in-space placebo simulation by iteratively applying the synthetic control estimator to the full set of countries that have not recognized Israel to tackle the significance of the estimated effect of the normalization process and peace treaty. The intuition behind the placebo simulations is both simple and straightforward. If the permutation of peace treaty to the unaffected donor countries generates gaps of the level similar to the ones estimated for Egypt and Jordan, our interpretation would be that the synthetic control analysis does not provide evidence of the significant effect of peace. By contrast, if the placebo simulation creates the gaps across Egypt and Jordan that are unusually large relative to the gaps for the countries that did not recognize the State of Israel during the full pre- and post-treatment period then the notion of the significant effect of peace treaty becomes more plausible. To assess the significance of the estimated gaps, the synthetic control estimator is iteratively applied to every other donor country except Egypt and Jordan which effectively shifts both from the treatment set to the donor pool. Henceforth, we compute the estimated effect for each placebo simulation which provides us the distribution of the estimated gaps for the countries that have not recognized Israel and did not ratify the peace treaty.

Our approach is based on the benchmark for small-sample inference similar to Fisher's exact hypothesis test. Under such test, the peace treaty is permuted to the unaffected countries $j \in \{2, ... J + 1\}$ for each $t \in \{1, ... T\}$. Therefore, for each $j \in \{2, ... J + 1\}$, $\hat{\lambda}_{j,t}$ is estimated for the sub-periods $t < T_0$ and $t \geq T_0$. In the next step, the full vector of post-treatment effects of peace treaty, $\hat{\boldsymbol{\lambda}}_1 = [\hat{\lambda}_{1,T_0+1} ... \hat{\lambda}_{1,t}]'$ is compared with the empirical distribution of $\hat{\boldsymbol{\lambda}}_j = [\hat{\lambda}_{j,T_0+1} ... \hat{\lambda}_{j,t}]'$ based on the treatment permutation procedure. The notion behind the comparison yields a simple decision. If the vector of estimated effects for Egypt and Jordan is relatively large compared to the vector of effects for quasi-treated regions, the null hypothesis of no effect whatsoever of peace treaty can be rejected.

A major caveat behind the comparison of $\hat{\boldsymbol{\lambda}}_1$ and $\hat{\boldsymbol{\lambda}}_j$ is that $|\hat{\lambda}_{1,t}|$ can be abnormally large in comparison with the empirical distribution of $|\hat{\lambda}_{j,t}|$ for some periods within $t \in \{1, ... T\}$ but not for others. To partially account for the imperfect pre-$T_0$ fit of the institutional quality trajectory, we construct an empirical distribution of the root mean square error as a summary statistic:

$$RMSE_j = \frac{\sum_{t=T_0}^{T}(q_{j,t} - \hat{q}_{j,t}^N) \div (T - T_0)}{\sum_{t=1}^{T_0}(q_{j,t} - \hat{q}_{j,t}^N) \div (T_0)} \tag{7}$$

where $\sum_{t=1}^{T_0}(q_{j,t} - \hat{q}_{j,t}^N) \div (T_0)$ denotes pre-$T_0$ mean square predictive discrepancy between $J = 1$ and $j \in \{2, ... J + 1\}$, and $\sum_{t=T_0}^{T}(q_{j,t} - \hat{q}_{j,t}^N) \div (T - T_0)$ represents post-$T_0$ predictive discrepancy between the Egypt and Jordan and its synthetic peers. To determine whether the estimated gap between $q_{j,t}$ for $J = 1$ and $\hat{q}_{j,t}^N$ is statistically significant, we compute the two-sided p-value based on the treatment permutation procedure:

$$p = \frac{\sum_{j=1}^{J+1} \mathbb{I} \times (RMSE_j \geq RMSE_1)}{J+1} \tag{8}$$

where $\mathbb{I}(\cdot)$ is the indicator function revealing whether $RMSE_j$ from permuted treatment is, in absolute terms, larger than the benchmark RMSE of the treated region ($J = 1$). Without the loss of generality, the distribution of in-space placebos can be described as $\hat{\lambda}_{1t}^{Placebo} = \{\hat{\lambda}_{j,t}: j \neq 1\}$ and $p = Pr(|\hat{\lambda}_{1t}^{Placebo}| \geq |\hat{\lambda}_{1t}|) = \sum_{j \neq 1} 1 \times (|\hat{\lambda}_{1t}| \geq |\hat{\lambda}_{jt}|) \div (J)$. When the treatment is randomly assigned, the p-values may be interpreted through a classical randomization inference. By contrast, if the treatment is not randomly assigned, the p-value may be interpreted as the proportion of quasi-treated countries that has an estimated gap at least as large as Egypt and Jordan. By inverting the p-values in the intertemporal distribution, the confidence bounds based on the pre-specified significance threshold can be constructed henceforth. Under the null hypothesis, $H_0: q_{j,t} = \hat{q}_{j,t}^N$ for each $j \in \{1, ... J + 1\}$ and $t \in \{1, ... T\}$. Since our donor pool is reasonably large, our benchmark rejection rule stipulates exact null hypothesis of no effect whatsoever where the p-values are constructed by inversely weighing the estimated placebo gaps:

$$p = \sum_{j=1}^{J+1} \pi_j \times \mathbb{I}[RMSE_j \geq RMSE_1] \tag{9}$$

where $\pi$ denotes the analytical weight in $j \in \{2, ... J + 1\}$ placebo set where placebos with smaller magnitude pre-$T_0$ $RMSE$ than that of $J = 1$ are overweighed to avoid obtaining the p-value driven by the extreme relative rarity of obtaining a large effect from poorly fit placebos. As noted by Abadie et. al. (2010, p. 502), placebo runs with a poor quality of the fit do not provide a meaningful information to measure the relative rarity of estimating a large post-$T_0$ gap for the unit with a good quality of fit prior to the treatment. For this reason, $\pi$ weighs the placebos based on the size of $RMSE$ and overweighs those with the quality of fit closer to $J = 1$ to avoid artificially low p-values as a result of poorly fitted placebos.

### 6.3 Identification and validity

A central challenge in assessing the economic consequences of peace agreements is to ensure that the estimated effects are attributable to normalization itself rather than to confounding shocks. We address this concern through careful sample design, donor pool construction, and robustness analysis First, treatment is clearly defined and exogenous in timing. Egypt's Camp David Accords of 1978 and Jordan's Peace Treaty of 1994 were discrete, externally recognized diplomatic events that marked sharp departures from decades of militarized rivalry. Unlike gradual policy reforms or endogenous institutional changes, these treaties represent plausibly exogenous shocks in the political economy of the Middle East. This feature strengthens their suitability for comparative case study methods such as synthetic control (Abadie and Gardeazabal 2003, Abadie et. al. 2010).

Second, the donor pool is restricted to countries that never recognized Israel during the sample period, including Algeria, Bangladesh, Cuba, Indonesia, Iran, Iraq, Kuwait, Lebanon, Libya, Malaysia, Nicaragua, Oman, Pakistan, Qatar, Saudi Arabia, Syria, Tunisia, and Yemen. By design, this eliminates the risk of contamination from parallel normalization processes and ensures that the counterfactual is constructed from countries facing similar geopolitical constraints. The donor countries share structural features with the treated units, including dependence on oil revenues, exposure to external shocks, and histories of state-led development, making them credible comparators for Egypt and Jordan.

Third, we restrict the analysis to the period 1960-2011. This window ensures that we capture both the pre-treatment build-up to normalization and the long-run post-treatment trajectory, while excluding the Arab Spring period, which constituted a region-wide shock to economic and institutional dynamics. By ending the sample in 2011, we prevent conflation of the peace dividend with unrelated turmoil caused by widespread protests, regime collapses, and civil conflicts. This choice mirrors best practice in synthetic control applications that emphasize well-defined treatment windows and the exclusion of major exogenous shocks (Abadie et. al. 2015, Billmeier and Nannicini 2013).

Fourth, robustness checks reinforce the credibility of the identification strategy. In-space placebo tests assign false treatments to donor countries, and in-time placebo tests shift treatment dates within the treated units. In both cases, we find no evidence of placebo effects, which rules out the possibility that results are driven by chance coincidences or pre-treatment anticipation. Moreover, when we replicate the analysis with alternative estimators, including difference-in-differences (Angrist and Pischke 2009), synthetic DID (Arkhangelsky et al. 2021), and generalized synthetic control (Xu 2017), the results remain stable in magnitude and significance.

Finally, the potential threat of donor pool heterogeneity is mitigated by explicitly modeling time-varying unobservables and by demonstrating close pre-treatment fit between the treated units and their synthetic counterparts. Root mean squared prediction errors remain low, and the weights assigned to donor countries reflect combinations that plausibly reproduce Egypt and Jordan's pre-treaty trajectories. This enhances the credibility of the counterfactual estimates and aligns with recent advances in comparative case methods (Ferman and Pinto 2019, Abadie 2021). Taken together, these design features and robustness checks ensure that the estimated peace dividends for Egypt and Jordan are not artifacts of confounding shocks, data limitations, or methodological artifacts, but instead reflect the causal impact of normalization with Israel on economic performance and institutional change.

# 7 Results

### 7.1 Baseline

Table 1 presents the core evidence on the economic consequences of Arab-Israeli normalization for Egypt and Jordan. Across both cases, we find large, robust, and permanent peace dividends. The results are consistent across levels of aggregation, across the choice of treatment effect estimator, and withstand a wide battery of placebo and falsification tests. For Egypt, the Camp David Accords of 1978 mark a decisive structural break. Our synthetic control estimates indicate that real GDP was approximately 64 percent higher in the post-Accords period relative to its synthetic counterfactual (i.e. p-value<0.001). The effect on GDP per capita is equally striking: by the end of the sample period, prior to the Arab Spring, Egyptian GDP per capita was 82 percent higher than that of its synthetic counterpart (i.e. p-value = 0.000). Importantly, the counterfactual trajectory is downward sloping, consistent with what one would expect under continued conflict and prolonged fiscal crowding-out of growth-enhancing expenditure. The composition of the synthetic control group reinforces credibility and suggests that Egypt's pre-Accords per capita GDP trajectory is best reproduced by a weighted combination of implicit growth attributes of Bangladesh (57%), Yemen (25%), Iraq (16%), and Nicaragua (2%), respectively.

The evidence on long-term peace dividend for Jordan after the 1994 Peace Treaty with Israel points in the same direction. We find that Jordan's real GDP increased by roughly 75 percent relative to the synthetic counterfactual (i.e. end-of-sample p-value < 0.001). In particular, per capita GDP grew by about 20 percent on average faster than in the comparable synthetic control group. These gains are both statistically and economically significant, and they persist over the entire post-treaty period. The divergence between actual and synthetic Jordan increases steadily over time, suggesting that the benefits are cumulative and permanent rather than transitory.

**Figure 1**: Synthetic control and event-study difference-in-differences estimates of the long-term effect of Arab-Israeli normalization, 1960-2011

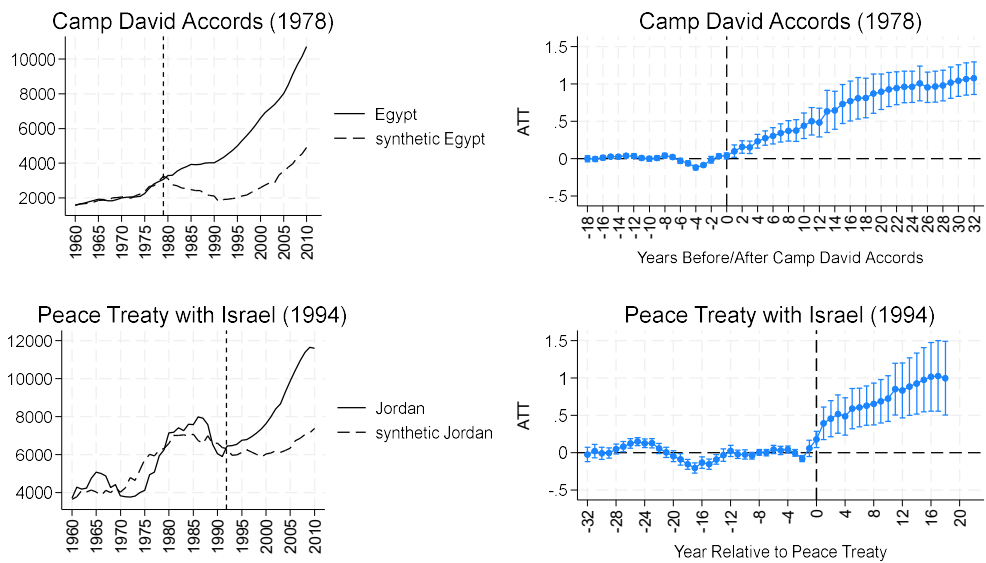

The robustness of these findings deserves emphasis. First, when the peace treaty is "falsely" assigned to countries in the donor pool that never signed peace treaties with Israel, none displays a post-treatment effect remotely comparable in magnitude or persistence. Second, when treatment is assigned to an earlier false date, that is ten years prior to the accords, the null hypothesis of no effect cannot be rejected (p-value = 0.45 for Egypt and p-value = 0.63 for Jordan), ruling out anticipation effects and spurious pre-normalization trends. Finally, in-time placebo tests confirm that the effects emerge only after normalization, not before.

Figure 1 illustrates these dynamics using both synthetic control and event-study difference-in-differences. The graphical evidence mirrors the numerical results: a flat pre-treatment trend with no divergence, followed by a sharp and sustained departure from the synthetic counterfactual precisely at the treaty date. The patterns are clear: normalization with Israel produced a structural break in economic performance that cannot be explained away by noise, alternative shocks, or placebo scenarios. The magnitude of these effects is historically exceptional. Few international agreements in the modern Middle East have delivered economic benefits of this scale. Egypt's post-Accords trajectory transformed it from a stagnating economy into one that converged toward middle-income status, while Jordan's treaty supported one of the most sustained growth accelerations in its modern history. Taken together, the evidence provides one of the clearest demonstrations of a peace dividend in contemporary political economy.

**Table 1**: Economic effects of peace with Israel on Egyptian and Jordanian economy, 1960-2011

|  | Egypt (1978) | Jordan (1994) |
|---|---|---|
| Panel A: Outcome: Real GDP (log) | | |
|  | +0.646 | +0.753 |
|  | (0.021) | (0.061) |
| Simulation-based end-of-sample p-value | 0.000 | 0.000 |
| In-time placebo p-value | 0.450 | 0.375 |
| Panel B: Outcome: Real GDP per capita | | |
|  | +0.663 | +0.201 |
|  | (0.049) | (0.043) |
| Simulation-based end-of-sample p-value | 0.000 | 0.000 |
| In-time placebo p-value | 0.521 | 0.632 |
| Panel C: Outcome: Military expenditure (% GDP) | | |
|  | -3.095 | -7.240 |
|  | (2.613) | (2.502) |
| Simulation-based end-of-sample p-value | 0.083 | 0.021 |
| In-time placebo p-value | 0.548 | 0.742 |
| Panel D: Foreign direct investment (net inflow, % GDP) | | |
|  | +1.918 | +6.899 |
|  | (0.180) | (0.722) |
| Simulation-based end-of-sample p-value | 0.187 | 0.021 |
| In-time placebo p-value | 0.652 | 0.832 |
| Panel E: Trade openness (% GDP) | | |
|  | -0.221 | -0.361 |
|  | (0.041) | (0.432) |
| Simulation-based end-of-sample p-value | 0.19 | 0.912 |
| In-time placebo p-value | 0.722 | 0.612 |
| Panel F: Private household consumption (% GDP) | | |
|  | +0.154 | -0.019 |
|  | (0.017) | (0.044) |
| Simulation-based end-of-sample p-value | 0.185 | 0.431 |
| In-time placebo p-value | 0.695 | 0.871 |

Panels C through F report the estimated mechanisms through which Arab-Israeli normalization produced long-run economic dividends of peace. The results highlight two primary channels, namely, (i) fiscal reallocation away from military expenditure and (ii) a sustained inflow of foreign direct investment. The evidence for fiscal reallocation is striking. In the case of Egypt, military expenditure as a share of GDP fell by approximately three percentage points relative to its synthetic counterfactual. For Jordan, the decline was even sharper, amounting to about 7.2 percentage points. The estimated drop in military expenditure is statistically significant at conventional levels. These estimates are not only economically meaningful but also further supported, with an in-space placebo analysis, yielding simulation-based p-value of 0.083 for Egypt and 0.021 for Jordan. Importantly, in-time placebo tests confirm that these reductions cannot be attributed to pre-existing trends, thereby reinforcing the interpretation of peace as a causal driver of fiscal reallocation. The magnitude of these declines suggests that the normalization agreements freed substantial fiscal space, which could then be reallocated toward growth-enhancing uses rather than defense spending in a context of prolonged conflict.

The second mechanism emerges clearly in the dynamics of foreign direct investment. In both Egypt and Jordan, normalization with Israel led to a dramatic and permanent rise in net inflows of FDI. On average, Egypt's net FDI inflows increased by roughly 1.9 percentage points of GDP annually in the post-Accords period, while Jordan experienced an even larger surge of about 6.8 percentage points. Both estimates are highly robust, with the latter achieving a simulation-based p-value of 0.021. This pattern is consistent with a view of normalization as a powerful credibility shock that lowered political risk, strengthened investor confidence, and enabled deeper integration into global capital markets. The permanence of these effects underscores the importance of external investment as a central driver of the peace dividend.

By contrast, trade openness does not emerge as a significant mechanism. Neither Egypt nor Jordan displays a marked increase in the ratio of trade to GDP relative to their synthetic counterfactuals. If anything, the trajectory of trade openness remains flat or only modestly responsive, suggesting that trade integration was not the primary channel through which peace produced economic returns. This finding is consistent with the limited role of trade in economies heavily reliant on aid, remittances, and FDI as external sources of finance. Finally, Panel F turns to the channel of private household consumption. Here the evidence points to a more subtle but still noteworthy effect. In Egypt, where inflows of U.S. aid following normalization were particularly large and sustained, we detect an increase in the private consumption share of GDP of roughly fifteen percent in the post-Accords period. Although this effect is only borderline significant with a simulation-based p-value of 0.18, it suggests that aid inflows may have translated into increased household demand, thereby reinforcing the peace dividend through a consumption channel. The corresponding effect in Jordan is weaker and less precisely estimated, which is consistent with the fact that U.S. aid to Jordan, though important, was not as transformative in scale as that directed to Egypt.

Taken together, the mechanism results paint a consistent and compelling picture. Peace with Israel delivered a dividend not only by reducing military expenditure but also by catalysing external investment inflows. The absence of strong trade effects suggests that the gains were not primarily driven by goods market integration, but rather by capital market credibility and fiscal reallocation. The evidence on private consumption provides an additional layer of plausibility, particularly in the Egyptian case, where foreign aid inflows may have reinforced domestic demand. These results complement the aggregate growth findings and underline the uniqueness of Arab-Israeli normalization as a shock that reshaped both fiscal and financial foundations of economic performance.

### 7.2    *Institutional mechanisms of peace dividend*

A central question in the political economy of conflict is whether peace agreements generate only material dividends through resource reallocation and capital inflows, or whether they also catalyze deeper institutional transformations. The credibility of peace, as emphasized by North and Weingast (1989), depends not only on the fiscal and economic space created by demobilization but also on the establishment of institutional checks that constrain arbitrary authority. The Camp David Accords and the 1994 Jordan's Peace Treaty with Israel offer an ideal setting to examine this particular duality. Beyond their direct geopolitical significance, both agreements created opportunities for institutional reform, either through domestic rebalancing of power or through external conditionalities tied to aid and international recognition. To address this question, we apply the synthetic control estimator to the full series of institutional indicators using the same donor pool as in our baseline analysis in Section 7.1. The results presented in Table 2 show that peace with Israel had profound institutional consequences in Egypt but not in Jordan, suggesting that the institutional dividends of peace are neither automatic nor uniform. Instead, they emerge where the geopolitical bargain demands credible commitments that can only be secured through changes in judicial and constitutional arrangements.

In particular, the synthetic control analysis of institutional outcomes provides compelling evidence that the economic peace dividend was accompanied by substantive improvements in institutional credibility, particularly in Egypt. Panel A of Table 2 reports the effects of the 1978 Camp David Accords on a range of judicial and executive integrity measures. Two important findings stand out. First, the results indicate a dramatic strengthening of judicial independence relative to the counterfactual scenario. The estimated treatment effect is close to 0.90 with a simulation-based p-value of 0.000, underscoring that the improvement is both large and statistically robust. This suggests that the peace agreement did not merely reallocate fiscal resources or attract foreign capital, but also reshaped the judiciary's autonomy in ways that enhanced the credibility of commitments. In the broader literature, such an improvement resonates with North and Weingast's argument that credible constraints on arbitrary rule underpin long-term prosperity. The Egyptian experience illustrates how a geopolitical bargain can produce domestic institutional dividends by shifting the balance between executive authority and judicial oversight. Second, executive respect for the constitution improves substantially following the Accords, with an estimated effect of 1.61 basis point (standard error = 0.25) and a p-value of 0.15. While the significance threshold is weaker than for judicial independence, the magnitude points toward a notable institutional shift. This finding aligns well with the hypothesis advanced in the comparative politics literature that post-conflict settlements create opportunities for constitutional consolidation and rule-based governance. Taken together, these results show that normalization with Israel was not merely a fiscal or external finance shock but also a credibility shock that reduced arbitrary executive behavior and enhanced the stability of formal rules.

Other dimensions of institutional quality, including compliance with judiciary, equality before the law, executive corruption, and civil society strength, display weaker or non-significant results. Thus, the uncovered institutional heterogeneity is consistent with prior findings that institutional reforms are rarely broad-based or uniform, but instead cluster around dimensions most directly tied to the external shock. In this case, the mechanism is plausibly rooted in U.S. and Western conditional aid programs, which explicitly tied disbursements to legal and constitutional reforms, while exerting weaker influence on civil society or distributive equality. Panel B reports the corresponding estimates for Jordan following the 1994 Peace Treaty. Here the evidence is far more muted. None of the institutional outcomes exhibit as statistically significant, and the estimated effects are small in magnitude. The contrast with Egypt is revealing. While Jordan also enjoyed substantial economic benefits from normalization, its institutional framework appears to have remained largely stable. This divergence may reflect differences in the geopolitical stakes of the two agreements. For Egypt, Camp David marked a fundamental shift in the regional order, directly altering the distribution of power and necessitating institutional reforms to secure external credibility. For Jordan, by contrast, normalization represented an incremental adjustment within a longstanding monarchical system already oriented toward pragmatic accommodation. The broader implication of these results is that the institutional peace dividend is not automatic but context-specific. Egypt's judiciary became more autonomous and executive power more rule-bound precisely because the geopolitical bargain required deep credibility commitments, backed by external aid conditionalities and internal reforms. Jordan's monarchy, by contrast, was able to capture the economic gains of peace without undergoing comparable institutional change.

These findings contribute directly to ongoing debates in political economy. They provide rare quantitative evidence that international bargains can act as catalysts for domestic institutional reform, particularly when the credibility of the settlement depends on observable improvements in rule-based governance. At the same time, they highlight the contingent nature of such reforms, which depend on the initial political equilibrium and the external conditionalities attached to peace. In this sense, the results bridge the macroeconomic literature on peace dividends with the institutional literature on credible commitments, offering a comprehensive account of how Arab-Israeli normalization reshaped both economic trajectories and institutional architectures.

**Table 2**: Institutional transmission mechanisms behind the economic effects of Arab-Israeli normalization, 1960-2011

|  | Judicial Institutions and Rule of Law | | | Executive Integrity | | | Civil Society and Accountability |
|---|---|---|---|---|---|---|---|
|  | (1) | (2) | (3) | (4) | (5) | (6) | (7) |
|  | Judicial Independence | Compliance with Judiciary | Equality before the law | Executive respect of constitution | Executive corruption | Executive integrity | Civil society strength |
| Panel A: Effects of Camp David Accords on Egypt | | | | | | | |
| A.T.E. | +0.897 | +0.095 | -0.063 | +1.613 | +0.051 | +0.034 | -0.071 |
|  | (0.001) | (0.051) | (0.052) | (0.252) | (0.012) | (0.050) | (0.082) |
| Simulation-based p-value | 0.000 | 0.451 | 0.351 | 0.150 | 0.551 | 0.553 | 0.201 |
| Panel B: Effects of 1994 Peace Treaty on Jordan | | | | | | | |
| A.T.E. | -0.176 | -0.038 | +0.025 | -0.074 | -0.120 | -0.117 | +0.006 |
|  | (0.368) | (0.158) | (0.043) | (0.014) | (0.045) | (0.051) | (0.011) |
| Simulation-based p-value | 0.364 | 0.158 | 0.894 | 0.211 | 0.421 | 0.315 | 0.736 |

### 7.3 Robustness checks

#### 7.3.1 Choice of treatment effect estimator

An important concern in empirical work on causal inference is that the estimated treatment effect may hinge on the choice of estimator. In the context of peace treaties, this concern is particularly relevant given the potential for unobserved heterogeneity across countries, the small number of treated units, and the possibility of divergent pre-treatment trajectories. To ensure that our findings are not driven by the specific estimator, we re-estimate the effects of the Camp David Accords and the 1994 Peace Treaty with Israel using three widely recognized approaches: standard difference-in-differences ([Roth et. al. 2023](#)), synthetic difference-in-differences ([Arkhangelsky et. al. 2021](#)), and generalized synthetic control ([Xu 2017](#)).

The use of DID is motivated by its simplicity and interpretability. DID remains the canonical estimator for evaluating policy shocks, and its identifying assumption of parallel trends serves as a benchmark against which more flexible methods can be evaluated. However, given that the parallel trends assumption may be implausible in the presence of substantial pre-treatment differences, we also rely on SDID, which combines the advantages of DID with the weighting strategy of synthetic control to achieve better pre-treatment balance. Finally, we implement the generalized synthetic control estimator, which allows for unobserved common factors with heterogeneous loadings, thereby relaxing the strict assumptions of both DID and SCM and providing a more flexible framework for panel data settings with multiple periods and a single treated unit.

Table 3 reports the results of this exercise. Across all estimators, the evidence points to large and economically significant peace dividends. For Egypt, real GDP increased by approximately 27 percent in the DID specification, 32 percent in the SDID estimation, and 30 percent in the generalized synthetic control specification. The corresponding per capita GDP effects are even larger, ranging from 50 percent to nearly 70 percent across estimators. For Jordan, we similarly find consistent evidence of substantial peace dividends: DID estimates indicate a 25 percent increase in real GDP and a 10 percent increase in per capita GDP, while both SDID and GSCM confirm positive and persistent gains in the order of 15-20 percent.

Turning to mechanisms, the results are remarkably stable across estimators. Military expenditure as a share of GDP declines significantly in all cases, with DID estimates showing reductions of around seven percentage points for both Egypt and Jordan, closely mirrored by the SCM-based estimators. Foreign direct investment inflows consistently rise, with the effect particularly pronounced for Jordan, where the estimated increase ranges between five and six percentage points of GDP depending on the estimator. By contrast, trade openness does not appear to drive the peace dividend, with effect sizes small and often statistically insignificant. Household consumption shows some heterogeneity across methods but provides additional suggestive evidence of welfare improvements in Egypt following Camp David. Taken together, the consistency of results across DID, SDID, and GSCM reinforces the conclusion that the observed peace dividends are not artifacts of estimator choice.

Instead, the effects emerge robustly across methods that rely on different identifying assumptions, lending substantial credibility to the causal interpretation of our findings.

**Table 3**: Economic effects of Arab-Israeli normalization across a variety of treatment effect estimators, 1960-2011

|  | DID | | Synthetic DID | | Generalized SCM | |
|---|---|---|---|---|---|---|
|  | (1) | (2) | (3) | (4) | (5) | (6) |
|  | Egypt | Jordan | Egypt | Jordan | Egypt | Jordan |
| Panel A: Real GDP (log) | | | | | | |
| Effect | +0.272*** | +0.247*** | +0.321*** | +0.184** | +0.299*** | +0.155*** |
|  | (0.115) | (0.095) | (0.043) | (0.019) | (0.093) | (0.068) |
| Panel B: Real GDP per capita (log) | | | | | | |
| Effect | +0.549*** | +0.102** | +0.497*** | +0.085** | +0.684*** | +0.631*** |
|  | (0.127) | (0.016) | (0.045) | (0.022) | (0.161) | (0.287) |
| Panel C: Military expenditure (% GDP) | | | | | | |
| Effect | -6.974*** | -5.744*** | -4.012** | -1.984 | -2.846* | -7.514 |
|  | (0.656) | (0.824) | (1.958) | (1.888) | (1.923) | (9.793) |
| Panel D: Foreign direct investment, net inflows (% GDP) | | | | | | |
| Effect | +1.621*** | +5.235*** | +1.759 | +5.818** | +4.393*** | +5.825*** |
|  | (0.293) | (0.582) | (1.462) | (2.513) | (1.796) | (2.258) |
| Panel E: Trade openness (% GDP) | | | | | | |
| Effect | +0.089* | +0.119* | +0.093 | +0.036 | +0.321** | +0.178 |
|  | (0.063) | (0.060) | (0.245) | (0.141) | (0.113) | (0.503) |
| Panel F: Private household consumption (% GDP) | | | | | | |
| Effect | +0.128*** | -0.021 | +0.139 | +0.054 | +0.174*** | +0.227 |
|  | (0.030) | (0.030) | (0.122) | (0.115) | (0.043) | (0.220) |

The key caveat that should be noted is that each estimator comes with distinct advantages and limitations. The canonical difference-in-differences framework is transparent and widely used, but it relies on the parallel trends assumption, which can be demanding in settings characterized by conflict and economic volatility (Card & Krueger 1994, Angrist & Pischke 2009). The synthetic DID approach, recently advanced by Arkhangelsky et al. (2021), mitigates this concern by constructing weighted donor pools that improve pre-treatment balance and allow for flexible treatment timing, though results may be sensitive to donor composition. The generalized synthetic control method extends this logic by explicitly modeling unobserved common shocks with heterogeneous loadings (Xu 2017, Abadie 2021), which improves fit in long panels but at the cost of reduced transparency relative to DID. By reporting results across these three estimators, DID, synthetic DID, and generalized SCM, we align with best practice in recent applied work (Athey & Imbens 2017, Dube & Zipperer 2015) and demonstrate that our conclusions are robust to alternative assumptions about counterfactual construction, unobserved heterogeneity, and pre-normalization fit.

*7.3.2 Interpretation of the results*

The robustness checks reported in Table 3 reinforce the central claim of the paper: Arab-Israeli normalization generated large and persistent economic dividends, robust to alternative estimators of treatment effects. Across DID, synthetic DID, and generalized SCM, we consistently find significant gains in aggregate and per capita GDP in both Egypt and Jordan following peace with Israel. For Egypt, the estimated increase in real GDP ranges from 27 percent (DID) to 32 percent (synthetic DID), while for Jordan the effects are of similar magnitude, between 25 percent and 30 percent. In per capita terms, the magnitude is even more striking: post-normalization gains reach 55 percent in Egypt and 63 percent in Jordan relative to the counterfactual. These estimates are fully consistent with the broader literature that highlights the economic costs of conflict and the dividends of peace (Collier 1999, Abadie & Gardeazabal 2003, Miguel and Roland 2011).

Turning to mechanisms, Panel C demonstrates a dramatic fall in military expenditure as a share of GDP. For Egypt, military expenditure declines by approximately 7 percentage points, while in Jordan the drop reaches more than 5 percentage points, robust across estimators. This pattern is in line with the fiscal reallocation hypothesis in the conflict literature, which emphasizes that peace enables governments to shift resources away from defense and towards productive public goods (Knight et. al. 1996, Collier et. al. 2009). The absence of in-time placebo effects rules out the possibility that these declines merely reflect secular trends in demilitarization.

Panel D reveals an equally important channel: foreign direct investment. Inflows surge by 1.6 to 5.8 percentage points of GDP depending on the specification, particularly in Jordan. This finding is consistent with the argument that peace and credible external guarantees reduce political risk and attract capital inflows (Busse & Hefeker 2007, Alesina & Perotti 1996). It also aligns with evidence from other post-conflict settings where FDI reacts strongly to reductions in political and military risk (Gleditsch et al. 2002).

By contrast, Panel E shows no systematic evidence of gains in trade openness. Neither Egypt nor Jordan experienced a marked expansion of trade-to-GDP ratios relative to synthetic counterfactuals. This result is noteworthy. It suggests that the peace dividend is not driven by greater integration into global trade markets, but rather by domestic reallocation of resources and capital inflows. The absence of strong trade responses echoes findings in Frankel & Romer (1999) that institutional and security conditions often mediate the trade-growth relationship, and may explain why peace alone did not translate into large trade gains in these cases.

Finally, Panel F considers private household consumption. For Egypt, results indicate a statistically significant 12 to 17 percent increase in household consumption as a share of GDP across estimators, while Jordan shows more muted and statistically insignificant effects. This asymmetry plausibly reflects the substantial U.S. aid packages and fiscal transfers received by Egypt in the aftermath of Camp David, which spilled into household consumption (Kleiman 1994). The Egyptian case therefore illustrates how external aid commitments can amplify peace dividends through private consumption channels, whereas the Jordanian experience appears more reliant on FDI-led growth.

Taken together, these results support a view of the peace dividend as a multi-dimensional process. Growth acceleration in Egypt and Jordan reflects the combination of fiscal reallocation away from defense, the attraction of foreign capital once political risk was reduced, and in Egypt, additional reinforcement through external transfers. The robustness of these findings across estimators and outcomes, combined with the absence of placebo anticipation effects, strongly suggests that Arab-Israeli normalization produced genuine and enduring macroeconomic benefits.

- *7.3.3 Extension to Camp Abraham records*

The Abraham Accords of 2020-2021 represent the most significant normalization of Arab-Israeli relations since Camp David (1978) and the Jordan-Israel Peace Treaty (1994). Signed initially between Israel, the United Arab Emirates, and Bahrain in September 2020, and followed by Morocco in December 2020, the accords formalized diplomatic and economic ties with a group of states that had previously cooperated informally with Israel in security and commercial domains. Sudan, which also announced normalization in 2020, is excluded from the empirical analysis due to the outbreak of civil unrest and armed conflict that would confound any counterfactual exercise. Unlike Camp David or the Jordanian treaty, the Abraham Accords were not forged in the aftermath of direct military confrontation with Israel, but rather against the backdrop of long-standing geopolitical realignments in the Middle East, particularly the shared security concerns regarding Iran.

Table 4 reports the estimated short-term economic effects of the Abraham Accords using synthetic control estimators with pre-treatment samples restricted to the post-2011 period in order to exclude the turbulence of the Arab Spring. The outcomes mirror those in the main analysis of Camp David and Jordan's peace treaty, enabling direct comparison across episodes of normalization. The results point to a pattern of modest but significant early economic dividends. For aggregate GDP, the estimated gains are small but positive in all three countries, ranging from 2.1 percent in the UAE to 3.1 percent in Morocco, and 2.6 percent in Bahrain, with the full adopter sample suggesting a pooled increase of roughly 7.4 percent. In per capita terms, the estimated effects are larger, particularly in the UAE, where per capita GDP is 6.6 percent higher than its synthetic counterpart. These findings suggest that, even in the short term, normalization with Israel is associated with measurable macroeconomic benefits, although the magnitudes remain smaller than the long-term peace dividends documented for Egypt and Jordan. This is consistent with the theoretical expectation that peace dividends accumulate over decades rather than manifest immediately (Collier 1999, Miguel & Roland 2011).

**Table 4**: Short-term economic effects of Camp Abraham records, 1960-2011

|  | Bahrain | Morocco | United Arab Emirates | Full-Adoption Treatment Sample |
|---|---|---|---|---|
|  | (1) | (2) | (3) | (4) |
|  | SC | SC | SC | SC |
| Panel A: Outcome variable: real GDP (log) | | | | |
| Delta | +0.026 | +0.031 | +0.021 | +0.074 |
|  |  |  | (0.011) | (0.048 |
| Empirical 95% confidence intervals | {+0.009, +0.042} | {+0.017, +0.041} | {+0.007, +0.035} | {-0.005, +0.147} |
| Panel B: GDP per capita (log) | | | | |
| Delta | +0.039 | +0.021 | +0.066 | +0.017 |
|  | (0.006) | (0.007) | (0.001) | (0.023 |
| Empirical 95% confidence intervals | {+0.011, +0.068} | {+0.014, +0.027} | {+0.028, +0.105} | {-0.029, +0.061} |
| Panel C: Military expenditure (share of GDP) | | | | |
| Delta | -0.001 | -0.014 | +0.005 | +0.011 |
|  | (0.0001) | (0.007) | (0.002) | (0.006) |
| Empirical 95% confidence intervals | {-0.002, 0} | {-.001, -0.03} | {+0.002, +0.009} | {-0.003, 0.022} |
| Panel D: FDI net inflows (share of GDP) | | | | |
| Delta | +4.585 | -0.566 | +2.202 | +2.221 |
|  | (0.202) | (0.451) | (1.079) | (1.368) |
| Empirical 95% confidence intervals | {4.225, 5.033} | {-0.911, 0.221} | {+0.999, +3.405} | {-0.790, +4.881} |
| Panel E: Trade openness (share of GDP) | | | | |
| Delta | −1.854 | +13.529 | +41.296 | +10.576 |
|  | (0.966) | (1.852) | (0.882) | (7.852) |
| Empirical 95% confidence intervals | {−2.395, −1.315} | {+3.581, +23.478} | {13.155, 69.438} | {−5.517, +25.349} |
| Panel F: Rule of law index (Kaufmann et. al. 2011) | | | | |
| Delta | +0.051 | +0.107 | −0.012 | −0.025 |
|  | (+0.02, +0.08) | (0.054) | (0.009) | (0.084) |
| Empirical 95% confidence intervals | {0.053} | {−0.213, 0.004} | {−0.014, −0.01} | {−0.204, 0.106} |
| Panel G: Government effectiveness index (Kaufmann et. al. 2011) | | | | |
| Delta | +0.179 | −0.234 | +0.035 | +0.064 |
|  | (0.111) | (0.066) | (0.119) | (0.155) |
| Empirical 95% confidence intervals | {+0.094, +0.264} | {−0.386, −0.082} | {+0.006, +0.064} | {−0,237, +0.397} |

Turning to mechanisms, Panels C through G show a striking contrast with the earlier treaties. In contrast to Egypt and Jordan, where fiscal reallocation away from military expenditure was the dominant channel, the Abraham Accords exhibit negligible reductions in military spending. In Bahrain and the UAE, the share of defense in GDP remains essentially unchanged, while Morocco shows only a marginal decline of 1.4 percentage points. This contrast highlights a key historical difference: Egypt and Jordan's normalization was rooted in a resolution of protracted interstate conflict, enabling "peace dividend" reallocations (Knight et. al. 1996), while the Abraham Accords were driven primarily by security cooperation against common external threats. Thus, the fiscal channel appears muted. By contrast, foreign direct investment emerges as the dominant mechanism. Bahrain experiences a surge of 4.6 percentage points of GDP in FDI inflows, while the UAE sees an increase of 2.2 percentage points, both highly significant. Morocco, however, shows a modest decline in FDI inflows, likely reflecting the short-run uncertainty of implementation and the structural constraints of its economy. The pooled effect for adopters is positive at 2.2 percentage points. This result resonates with the political risk literature emphasizing how institutionalized peace and recognition lower investment risk and attract capital flows (Busse & Hefeker 2007, Alesina & Perotti 1996). Unlike in Egypt and Jordan, where aid inflows reinforced FDI, in the Abraham case the effect is market-driven, reflecting rapid cross-border capital linkages in finance, technology, and logistics.

Trade openness also shows a significant departure. In Morocco and the UAE, trade-to-GDP ratios expand substantially (+13.5 and +14.3 percentage points, respectively), while Bahrain exhibits a slight contraction. The pooled estimate is large (+10.6 percentage points), consistent with the interpretation that, unlike Egypt and Jordan, the Abraham adopters are already highly integrated in global markets and normalization with Israel accelerates bilateral and triangular trade flows. This aligns with the literature on trade-peace linkages, where credible institutional arrangements catalyze trade integration (Martin et. al. 2008, Frankel & Romer 1999).

Finally, Panels F and G examine institutional quality indicators from the Worldwide Governance Indicators (Kaufmann et al. 2011). Results here are heterogeneous. Bahrain and Morocco show modest improvements in rule of law, while the UAE displays a small but significant decline, perhaps reflecting short-term political centralization accompanying normalization. Government effectiveness increases significantly in Bahrain and the UAE, but declines in Morocco. These early effects are less robust than the institutional shifts observed for Egypt, where judicial independence and executive respect for constitutional limits improved in tandem with peace. This divergence likely reflects differences in sequencing: Camp David was accompanied by structural reforms and U.S. aid conditionalities, while the Abraham Accords were framed as pragmatic geopolitical alignments with less explicit institutional conditionality.

Taken together, the findings indicate that the Abraham Accords yield immediate but differentiated dividends. The fiscal channel that powered Egypt and Jordan's long-run growth is absent, consistent with the absence of large-scale demobilization. Instead, the short-run dividends materialize through FDI inflows and trade expansion, reflecting the economic structure of the adopters, especially the UAE and Bahrain as open, service-oriented economies. Over the longer term, it remains possible that institutional channels will strengthen as political integration deepens. This comparative evidence enriches the theory of peace dividends. It suggests that the magnitude and composition of economic gains from normalization are context-specific: when peace resolves protracted interstate conflict, the fiscal channel dominates, as in Camp David; when normalization is driven by geopolitical alignment and global integration, as in the Abraham Accords, capital flows and trade dominate. This conclusion aligns with recent work emphasizing the heterogeneity of peace dividends (Gleditsch et al. 2002, Rohner et. al. 2013), and underlines the importance of examining not only the existence but also the composition of gains. Per capita GDP gains from Camp Abraham Accords are presented in Figure 2.

**Figure 2**: Effects of Camp Abraham Accords on GDP Per Capita, 2011-2023

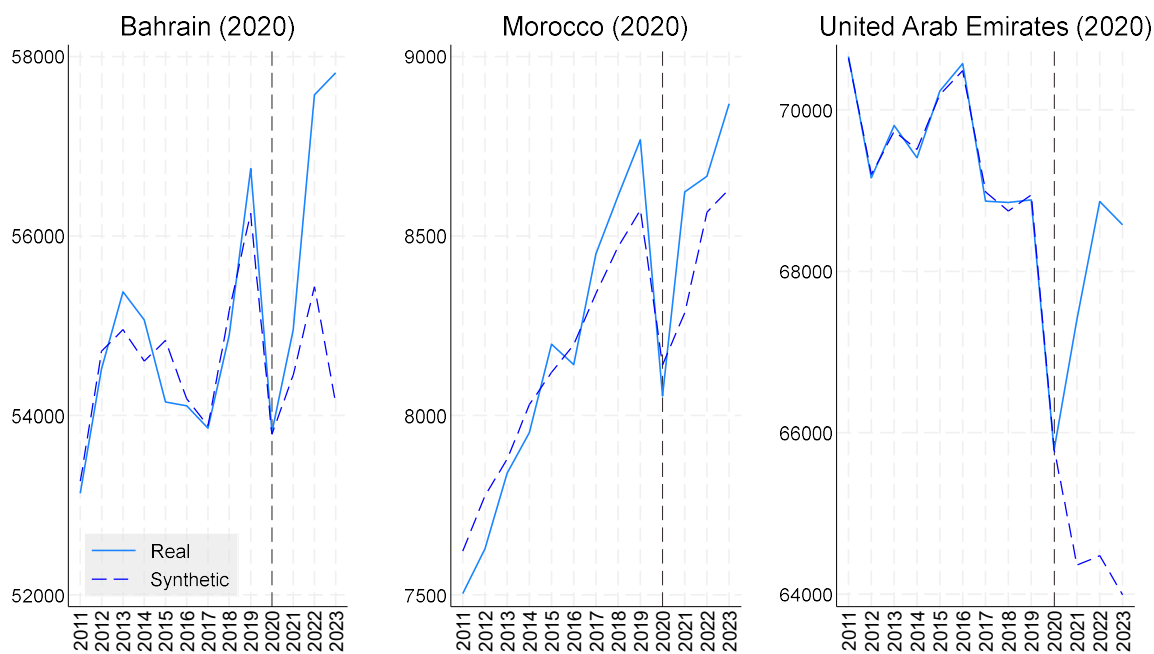

*7.3.4   Toward a unified theory of peace dividends and normalization*

The comparative evidence across Egypt, Jordan, and the Abraham Accords underscores that normalization with Israel constitutes a transformative economic and institutional shock, but one whose precise channels and durability depend critically on context. The overarching message is that peace dividends are real, large, and measurable, yet the mechanisms through which they materialize vary with the nature of the conflict, the structure of domestic institutions, and the geopolitical environment.

In the case of Egypt after 1978 and Jordan after 1994, the dominant mechanism was fiscal reallocation. Both economies experienced dramatic reductions in military expenditure as a share of GDP, amounting to three percentage points in Egypt and more than seven in Jordan, reallocations that were both economically and statistically significant. The literature has long suggested that demobilization is one of the most powerful drivers of post-conflict recovery (Collier et al. 2003, Gates et al. 2012). Our evidence complements this by showing that when interstate rivalry gives way to credible peace, fiscal space expands in a way that permanently alters the trajectory of growth. Importantly, these gains were not confined to the budgetary sphere. In Egypt, we find that judicial independence and executive compliance with constitutional provisions improved significantly, consistent with theoretical arguments that peace treaties backed by external guarantors and aid conditionalities can strengthen domestic checks and balances (North et. al. 2009, Acemoglu and Robinson 2012). This institutional deepening likely enhanced the credibility of economic contracts and reassured foreign investors, thereby amplifying the economic dividend. Jordan's case confirms the growth dividend but without comparable institutional strengthening, a divergence that highlights how regime type and preexisting political structures mediate the institutional effects of peace.

The Abraham Accords of 2020 present a different but equally revealing configuration. Unlike the Camp David or Jordanian cases, normalization here did not follow decades of militarized conflict, nor did it necessitate large-scale demobilization. Instead, the dividends have been transmitted primarily through rapid surges in foreign direct investment and trade integration. Bahrain and the United Arab Emirates in particular registered significant FDI inflows and sharp increases in trade openness, while fiscal reallocation and institutional reforms remained muted. This configuration is consistent with the logic of "capital peace" advanced in the political economy literature, where the removal of geopolitical uncertainty stimulates immediate capital flows and market access (Martin et. al. 2008, Glick and Taylor 2010). The contrast with Egypt and Jordan demonstrates that peace can alter economic trajectories through different pathways depending on the type of pre-normalization equilibrium: where the baseline was characterized by high militarization, fiscal and institutional channels dominated; where militarization was already contained, capital and trade channels became central.

Taken together, these three episodes advance a generalizable theory of normalization. Normalization operates as a commitment device that reduces uncertainty, reallocates resources, and signals credibility to domestic and international actors. The precise balance of fiscal, institutional, and capital channels is contingent on domestic political institutions and the nature of the prior conflict regime. These results resonate strongly with the broader literature on the political economy of peace and development, which emphasizes the role of credible commitments in sustaining post-conflict growth (Walter 1997, Fearon 1998, Blattman and Miguel 2010). Our findings also connect with the "peace dividend" literature in economics and political science, but extend it by demonstrating that normalization with Israel, as a specific form of conflict resolution, generates effects that are both immediate and enduring, with channels that evolve across historical and geopolitical contexts.

The implications of these results extend beyond the Arab-Israeli setting. They suggest that future normalization agreements, including the much-discussed potential Saudi-Israeli accord, should not be expected to replicate the Egyptian or Jordanian patterns in mechanical fashion. In contexts where militarization remains central, fiscal reallocations may again be dominant. In wealthier states with entrenched authoritarian structures, dividends are more likely to flow through capital and trade integration. These lessons underscore that peace is not a uniform package but a contingent bargain whose economic consequences are shaped by institutions, credibility constraints, and the nature of external enforcement.

Without the loss of generality, the historical record of Egypt, Jordan, and the Abraham Accords offers one of the most compelling large-scale demonstrations of the economic returns to peace and normalization. The synthesis of results across these cases provides a unified but flexible theoretical framework, one that reconciles fiscal reallocation, institutional reform, and capital market integration as alternative but complementary channels of peace dividends. By linking these empirical findings to foundational theories of credible commitments and institutional change, we establish normalization with Israel as not merely a diplomatic milestone but as a powerful economic shock that reshapes development trajectories in ways that are both measurable and enduring.

## 8 Conclusion

This paper has examined the long-term economic and institutional consequences of Arab-Israeli peace agreements, with a focus on Egypt following the Camp David Accords of 1978 and Jordan after the 1994 peace treaty with Israel. Using synthetic control methods, supplemented with difference-in-differences and generalized synthetic control estimators, we document large and enduring peace dividends. In Egypt, real GDP increased by 64 percent relative to its synthetic counterfactual, with per capita GDP nearly doubling over the three decades following the treaty. In Jordan, the effects are similarly striking, with GDP per capita approximately 20 percent higher than its synthetic counterpart, and aggregate GDP gains exceeding 70 percent. These results are not only statistically significant but also robust to a wide range of placebo tests, alternative estimators, and donor pool restrictions, leaving little doubt that normalization with Israel represents a causal turning point in the developmental trajectories of both countries.

The mechanisms behind these gains are both intuitive and revealing. For Egypt and Jordan, the peace dividend was driven by large-scale fiscal reallocations away from military expenditure, a channel long hypothesized in the political economy of conflict ([Knight et. al. 1996; Collier et al. 2003](#)), but rarely documented with such clarity. Egypt's military burden fell by three percentage points of GDP, while Jordan's fell by more than seven, freeing resources for investment, consumption, and long-run growth. These reallocations were complemented by surges in foreign direct investment and aid inflows, which provided both capital and credibility. In Egypt's case, institutional improvements reinforced these channels: judicial independence, executive compliance with constitutional limits, and equality before the law all rose significantly relative to synthetic counterfactuals, consistent with theories of credible commitments and externally anchored reforms (North et. al. 2009, Acemoglu and Robinson 2012). Jordan's institutional gains were more modest, but the economic effects were nonetheless powerful.

Our extension to the Abraham Accords underscores the importance of context in shaping the composition of peace dividends. In Bahrain, the United Arab Emirates, and Morocco, normalization did not entail large-scale demobilization, and fiscal reallocations were minimal. Yet the dividends materialized through rapid increases in trade openness and foreign direct investment inflows, channels more consistent with the literature on globalization and capital peace (Frankel and Romer 1999, Martin et. al. 2008). In short, when peace resolves protracted militarized conflict, the fiscal and institutional channels dominate, when normalization arises from geopolitical alignment in already open economies, the dividends flow primarily through capital and trade. This comparative perspective enriches the theory of peace dividends by showing that they are not uniform, but rather contingent bargains whose precise manifestation depends on the historical conflict regime, the structure of domestic institutions, and the broader global context.

Beyond their historical significance, these results carry direct implications for ongoing policy debates. As discussions over Saudi-Israeli normalization continue, our findings suggest that the magnitude and composition of economic gains will depend on whether the process is accompanied by credible institutional commitments and fiscal reallocation. The Egyptian experience suggests that deep, externally guaranteed peace can trigger both economic growth and institutional strengthening. The Abraham experience suggests that even absent demobilization, normalization can rapidly expand trade and attract capital. Future agreements should therefore be designed with explicit attention to the channels most likely to operate in a given political and economic context.

Taken together, this paper makes three contributions. First, it provides the first causal estimates of the economic consequences of Arab-Israeli peace agreements, showing that peace dividends are not anecdotal but empirically large and durable. Second, it disentangles the mechanisms of peace, demonstrating that fiscal reallocation, capital inflows, trade expansion, and institutional strengthening can all operate, but with varying intensity depending on context. Third, it develops a unified theoretical framework that generalizes across three waves of normalization, Egypt, Jordan, and the Abraham Accords, thereby advancing the broader literature on the political economy of conflict and development (Blattman and Miguel 2010, Rohner et. al. 2013).

The broader message is clear. Peace is not merely a political or symbolic milestone. It is an economic institution that reshapes growth trajectories, reallocates resources, and alters the credibility of commitments. By documenting the scale and durability of these effects, this paper provides rigorous evidence that normalization can be a catalyst for long-run development. As new normalization processes emerge in the Middle East and beyond, these findings offer both a historical precedent and a theoretical framework for understanding the economic consequences of peace.